\documentclass[fleqn,usenatbib]{mnras}

\usepackage{newtxtext,newtxmath}

\usepackage[T1]{fontenc}

\DeclareRobustCommand{\VAN}[3]{#2}
\let\VANthebibliography\thebibliography
\def\thebibliography{\DeclareRobustCommand{\VAN}[3]{##3}\VANthebibliography}

\usepackage{graphicx}	
\usepackage{amsmath}	

\title[Dissociation of H$_2$O in WASP-76~b]{Revealing H$_2$O dissociation in WASP-76~b through combined high- and low-resolution transmission spectroscopy}

\author[Gandhi et al.]{
Siddharth Gandhi$^{1,2,3}$\thanks{E-mail: gandhi@strw.leidenuniv.nl},
Rico Landman$^{1}$,
Ignas Snellen$^{1}$,
Luis Welbanks$^{4,5}$,
Nikku Madhusudhan$^{6}$
\newauthor
and Matteo Brogi$^{7,8}$
\\
$^{1}$Leiden Observatory, Leiden University, Postbus 9513, 2300 RA Leiden, The Netherlands\\
$^{2}$Department of Physics, University of Warwick, Coventry CV4 7AL, UK\\
$^{3}$Centre for Exoplanets and Habitability, University of Warwick, Gibbet Hill Road, Coventry CV4 7AL, UK\\
$^{4}$School of Earth \& Space Exploration, Arizona State University, Tempe, AZ 85257, USA \\
$^{5}$NHFP Sagan Fellow\\
$^{6}$Institute of Astronomy, University of Cambridge, Madingley Road, Cambridge, CB3 0HA, UK\\
$^{7}$Dipartimento di Fisica, Universit\`a degli Studi di Torino, via Pietro Giuria 1, I-10125, Torino, Italy\\
$^{8}$INAF-Osservatorio Astrofisico di Torino, Via Osservatorio 20, I-10025 Pino Torinese, Italy\\
}

\date{Accepted XXX. Received YYY; in original form ZZZ}

\pubyear{2023}

\begin{document}
\label{firstpage}
\pagerange{\pageref{firstpage}--\pageref{lastpage}}
\maketitle

\begin{abstract}
Numerous chemical constraints have been possible for exoplanetary atmospheres thanks to high-resolution spectroscopy (HRS) from ground-based facilities as well as low-resolution spectroscopy (LRS) from space. These two techniques have complementary strengths, and hence combined HRS and LRS analyses have the potential for more accurate abundance constraints and increased sensitivity to trace species. In this work we retrieve the atmosphere of the ultra-hot Jupiter WASP-76~b, using high-resolution CARMENES/CAHA and low-resolution HST WFC3 and Spitzer observations of the primary eclipse. As such hot planets are expected to have a substantial fraction of H$_2$O dissociated, we conduct retrievals including both H$_2$O and OH. We explore two retrieval models, one with self-consistent treatment of H$_2$O dissociation and another where H$_2$O and OH are vertically-homogeneous. Both models constrain H$_2$O and OH, with H$_2$O primarily detected by LRS and OH through HRS, highlighting the strengths of each technique and demonstrating the need for combined retrievals to fully constrain chemical compositions. We see only a slight preference for the H$_2$O-dissociation model given that the photospheric constraints for both are very similar, indicating $\log(\mathrm{OH/H_2O}) = 0.7^{+0.3}_{-0.3}$ at 1.5~mbar, showing that the majority of the H$_2$O in the photosphere is dissociated. However, the bulk O/H and C/O ratios inferred from the models differs significantly, and highlights the challenge of constraining bulk compositions from photospheric abundances with strong vertical chemical gradients. Further observations with JWST and ground-based facilities may help shed more light on these processes.
\end{abstract}




\section{Introduction} \label{sec:intro}

Observations of ultra-hot Jupiters (UHJs) have yielded fascinating insights into the physics which occurs in their atmospheres. These planets orbit at very short distances from their host star and thus are strongly irradiated, with equilibrium temperatures exceeding 2000~K. Numerous UHJs have been characterised in transmission and emission geometries with a range of observations, in particular through ground-based high-resolution spectroscopy (HRS). These observations are at very high spectral resolution (R$\gtrsim25,000$) and hence allow us to resolve the individual lines of spectrally active species in emergent and transit spectra \citep[see e.g.,][for a review]{birkby2018}. Numerous chemical species have been detected with HRS, for example refractories such as Fe, Na and TiO in the optical \citep[e.g.,][]{hoeijmakers2019, ehrenreich2020, borsa2021, prinoth2022}, which are generally more prominent in transmission geometries but have also been detected in emission \citep[e.g.][]{kasper2022, ramkumar2023, johnson2023}. In addition, volatiles such as H$_2$O, CO, OH and HCN have also been detected in the infrared in both transmission and emission geometries \citep[e.g.,][]{birkby2013, hawker2018,landman2021, vansluijs2023, yan2023, finnerty2024}. In addition, retrievals of high resolution spectra have recently become possible thanks to likelihood based approaches, allowing for abundance estimates of the chemical species in the atmosphere \citep[e.g.,][]{brogi2019, gandhi2019_hydrah, gibson2020, pelletier2021}. These quantitative assessments are key to constrain their formation history \citep[e.g.,][]{oberg2011, mordasini2016}.

Space-based low-resolution spectroscopy (LRS) has also been used to extensively study the atmospheres of UHJs, in particular with Hubble Space Telescope's (HST) Wide Field Camera 3 (WFC3) instrument in the infrared. This probes the $\sim$1.1-1.7~$\mu$m range where H$_2$O has strong molecular opacity. The observations have shown that such planets often exhibit muted H$_2$O features \citep[e.g.,][]{sheppard2017, evans2017, arcangeli2018, mansfield2018, kreidberg2018}, thought to arise from thermal dissociation of H$_2$O into OH and H in the upper atmosphere at pressures $\lesssim10^{-1}$~bar due to the very high temperatures \citep[e.g.,][]{parmentier2018, lothringer2018}. Retrievals of such spectra have been ubiquitous in both transmission and emission geometries for over a decade, across a wide range of planetary masses and equilibrium temperature \citep[e.g.,][]{madhu2009, madhu2014_ret, kreidberg2014, line2016_ret, changeat2022}.

These two observational approaches have great complementarity. LRS is more sensitive to continuum absorption whereas HRS probes the cores of spectral lines generated at higher altitudes in the atmosphere \citep{gandhi2020_clouds}. Hence, given that many thousands of spectral lines are typically correlated for each molecular species, HRS also allows for unambiguous detections of species in the atmosphere. On the other hand, spectral features from LRS can often be clearly seen in the observations, but HRS datasets are often lower signal-to-noise per pixel and thus the atmospheric signatures must often be extracted through techniques such as cross-correlation against model spectral templates. Constraining H$_2$O in the atmospheres of exoplanets is also more challenging with HRS because of strong telluric absorption \citep[e.g.][]{brogi2012, lockwood2014, pelletier2021, webb2022}, and often means that wavelength ranges outside of the peak H$_2$O bands must be used to detect the species. However, both methods are capable of tight constraints on various species in the atmosphere, and have shown great promise in characterising the atmospheres of a wide range of exoplanets \citep{barstow2017, pinhas2019, welbanks2019, line2021, gibson2022, maguire2023, pelletier2023, gandhi2023}.

Combining HRS and LRS observations is therefore highly advantageous. \citet{brogi2019} showed that HRS and LRS observations may be combined into a retrieval framework, which crucially opened up the possibility of combining the strengths of each method. This was first demonstrated by retrievals of HST+Spitzer and CRIRES/VLT dayside spectra of the hot Jupiter HD~209458~b \citep{gandhi2019_hydrah}, showing that the combined approach allows for highly robust atmospheric constraints with higher significance detections and an increased sensitivity to trace species. \citet{kasper2022} used optical high-resolution spectra from MAROON-X/Gemini-N with infrared HST and Spitzer observations of the ultra-hot Jupiter MASCARA-2~b. The combination of such optical and infrared observations provided constraints on refractory-to-volatile ratios, important tracers of planetary formation \citep{lothringer2021}. More recently, \citet{boucher2023} combined SPIRou observations of the sub-Saturn WASP-127~b with HST+Spitzer to constrain H$_2$O and carbon-rich species in the atmosphere, and \citet{smith2024} combined JWST and IGRINS observations for the hot Jupiter WASP-77A~b.

In this work we present combined retrievals of low- and high-resolution observations of the primary eclipse of the ultra hot Jupiter WASP-76~b \citep{west2016}, using CARMENES \citep{quirrenbach2014, landman2021} and HST+Spitzer observations \citep{fu2021} to probe the dissociation of H$_2$O in the atmosphere. Primary eclipse spectra are sensitive to the atmospheric chemistry, and hence are ideal to explore chemical composition of exoplanets \citep[see e.g.,][]{madhu2019}. One of the key insights our combined approach allows is for comparison of the atmospheric constraints over a similar spectral range, as both the CARMENES and HST WFC3 range probe wavelengths in the $\sim$1.1-1.7~$\mu$m range. In particular, some species such as OH have been robustly detected with the HRS data given their strong spectral lines over the continuum \citep{landman2021, cheverall2023}, but others such as H$_2$O are more sensitive to LRS given the lack of telluric absorption obscuring the molecular absorption bands \citep{fu2021}. Furthermore, combined observations also probe a wider range in pressures, with HRS typically being more sensitive to lower pressures \citep{gandhi2020_clouds, hood2020}. At the temperatures of ultra-hot Jupiters such as WASP-76b, we expect H$_2$O to be thermally dissociated into O and OH, with dissociation onsetting at some characteristic pressure in the atmosphere. We therefore also explore a dissociation model that allows for a vertical chemical gradient for H$_2$O in the atmosphere, and compare and contrast the constraints we obtain from each model across the high-resolution, low-resolution, and combined datasets. 

In the next section we discuss the observations and analysis in more detail. This is followed by the retrieval setup as well as a description of the dissociation model which allows us to model the pressure dependent H$_2$O abundance in the atmosphere. We then discuss the results, comparing and contrasting the two models across the HRS only, LRS only and combined analyses. Finally, we present the conclusions of our work and the implications for combined retrievals going forwards. 

\section{Observations and Analysis}\label{sec:observations}

In this section we describe the high-resolution and low-resolution observations. We use primary eclipse observations of WASP-76~b from the CARMENES spectrograph on CAHA \citep{landman2021, sanchezlopez2022} as well as from the HST and Spitzer spectrographs \citep{fu2021}. These observations were chosen because of their high signal-to-noise, and have been used previously to explore the chemical composition of the atmosphere of WASP-76~b.

\subsection{CARMENES observations}
\subsubsection{Data cleaning}
\begin{figure}
    \centering
    \includegraphics[width=\linewidth]{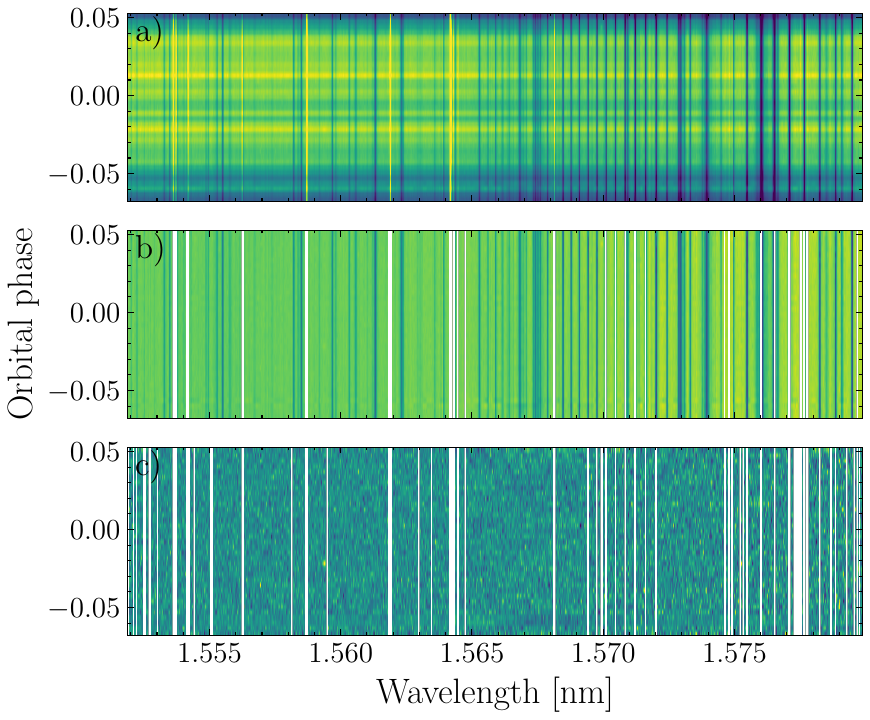}
    \caption{Visiualisation of the data cleaning steps applied to the CARMENES data for one spectral order showing a) the raw spectra, b) the spectra after removing the continuum and masking the tellurics, c) the residuals after applying SysRem to the data. Vertical white bars mark the masked spectral channels (Section 2.1).}
    \label{fig:data_analysis}
\end{figure}

The high-resolution observations were obtained during the 4 October 2018 transit of WASP-76b with the near-infrared channel of CARMENES (Calar Alto high-Resolution search for M dwarfs with Exoearths with Near-infrared and optical Echelle Spectrographs; \citet{quirrenbach2014}). The pipeline reduced data are publicly available in the CAHA archive, and have previously been analysed in \citet{landman2021}, \cite{casasayasbarris2021} and \cite{sanchezlopez2022}. Full details on the reduction steps are given in \citet{landman2021}, but we repeat the main steps here. First, we remove spectral orders 1, 8-11 and 17-22, which are all heavily dominated by telluric absorption. Next, we remove the continuum by convolving the spectra with a Gaussian kernel with a standard deviation of 500 pixels and dividing by this. After that, we mask the major telluric absorption and emission lines by removing wavelength bins that have on average less than 40\% flux or more than 10\% excess flux w.r.t the continuum. Subsequently, we mask any remaining 5$\sigma$ outliers to remove the effect of some spurious pixels. Finally, we use the SysRem algorithm \citep{tamuz2005} to detrend the data and remove the remaining tellurics. \citet{landman2021} studied the effect of the number of SysRem iterations on the cross-correlation signal and found that 9 iterations provided the highest signal-to-noise ratio detection. Hence, as we are using the same code, we also choose to apply 9 SysRem iterations to the data. This is relatively aggressive cleaning and could partially be the reason that we are unable to constrain the water abundance with the high-resolution data only (see section~\ref{sec:results}). However, our key goal is to ensure that the planetary parameters are unbiased, even if this means some of the planetary signal is lost during this detrending. Figure \ref{fig:data_analysis} visualises the data reduction steps for a single spectral order.

\subsubsection{Likelihood \& model reprocessing}\label{sec:hr_analysis}
Given the cleaned data, we use the log-likelihood mapping from \citet{gibson2020} with optimised noise scaling $\beta$, which uses the following likelihood:
\begin{equation}\label{eq:likelihood}
    \ln \mathcal{L}_\mathrm{HRS} = - \frac{N}{2} \ln \frac{\chi^2}{N},
\end{equation}
where $\chi^2$ is defined as:
\begin{equation}\label{eq:chi2}
    \chi^2 = \sum_i \frac{(f_i - m_i)^2}{\sigma_i^2}.
\end{equation}
Here $N$ is the number of data points, $f_i$ and $m_i$ the data and model at wavelength bin $i$ respectively, and $\sigma_i$ the uncertainties obtained from pipeline propagated through the data reduction steps. In contrast to \citet{gibson2020}, we do not include a scaling factor for the global strength of spectral lines in the model. The details of the modelling is discussed in section~\ref{sec:retrieval}. To avoid the data analysis steps biasing our retrievals, we apply the same cleaning steps to the generated models. First, we high-pass filter the generated model spectra in the same way as the data. Next, we apply the SysRem detrending to the generated model matrix. To speed this up, we use the fast model filtering technique from \citet{gibson2022}. Given the model matrix $\textbf{M}$, with its rows containing the modelled, high-pass filtered transit depth at the corresponding phase, and the SysRem matrix $\textbf{U}$, with its rows containing the 9 SysRem modes, the reprocessed model $\textbf{M'}$ is given by \citep{gibson2022}:
\begin{equation}
    \textbf{M'} = \textbf{M} - \textbf{U}(\mathbf{\Lambda} \textbf{U})^{\dagger} \mathbf{\Lambda}\textbf{M}.
\end{equation}
Here $\Lambda$ is a diagonal matrix containing the data uncertainties and $\dagger$ denotes the Moore-Penrose pseudo-inverse. For more details on this method, refer to \citet{gibson2022}. This reprocessed model is then used as the $m_i$ terms in the calculation of the $\chi^2$ (Eq. \ref{eq:chi2}).

\begin{figure*}
\centering
	\includegraphics[width=\textwidth,trim={0cm 0cm 0cm 0},clip]{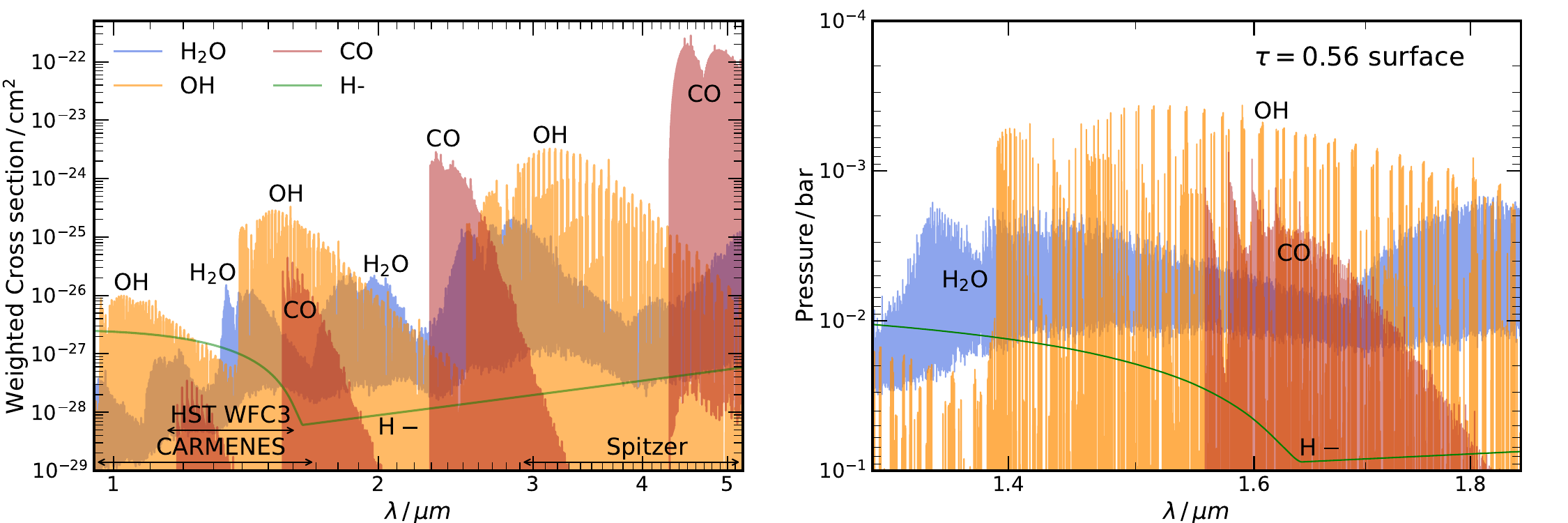}
    \caption{Left panel: Abundance weighted cross section of each species at a temperature of 2900~K and 1~mbar. We also show the wavelength ranges of the CARMENES, WFC3 and Spitzer observations. Right panel: Pressure level where the optical depth of the atmosphere reaches a values of 0.56 as a function of wavelength for each species, assuming thermal dissociation and a solar composition atmosphere. This is shown for the wavelength region with significant OH lines where the CARMENES and WFC3 observations overlap.}     
\label{fig:cs}
\end{figure*}

\subsection{HST and Spitzer observations}

The low resolution observations were obtained from \citet{fu2021}, GO Program ID: 14260 (PI: Deming) for the HST WFC3 data and GO Program ID: 13038 (PI:Stevenson) for the Spitzer data. These encompass one transit of WASP-76~b observed with the HST WFC3, with wavelengths between $\sim$1.1-1.6~$\mu$m, with 95 exposures with 104s exposure times for each. The HST observations showed a clear constraint for H$_2$O, probing the $\sim$1.4~$\mu$m H$_2$O feature. The Spitzer observations use the IRAC 3.6~$\mu$m and 4.5~$\mu$m channels, with two transits and one transit for each channel respectively. These showed excess absorption in the 4.5~$\mu$m channel, indicating the presence of CO. To evaluate the likelihood for the LRS observations, we use the $\chi^2$ value, given by 
\begin{align}
    \chi^2 &= \sum_i \frac{(f_i - m_i)^2}{\sigma_i^2},
\end{align}
where the sum $i$ is over the low-resolution data points. As in section~\ref{sec:hr_analysis}, $f_i$ and $m_i$ refer to the data and model respectively, and the error on each data point is given by $\sigma_i$. We evaluate the likelihood through
\begin{align}
    \ln \mathcal{L}_\mathrm{LRS} &= - \frac{1}{2} \chi^2.
\end{align}

\subsection{Combining the observations}

The total log likelihood for the combination of HRS and LRS observations is thus
\begin{align}
    \ln \mathcal{L}_\mathrm{tot} = \ln \mathcal{L}_\mathrm{HRS} + \ln \mathcal{L}_\mathrm{LRS}.
\end{align}
This combination of the likelihoods for the high-resolution and low-resolution observations is what allows us to perform retrievals on the datasets simultaneously. Importantly, $\ln \mathcal{L}_\mathrm{tot}$ is automatically weighted by the data quality, poor constraints for one dataset will influence the total value of the likelihood to a lower extent over data with higher precision. This also means that despite the significantly higher number of data points for the HRS data, the generally larger error bars will counter-balance this in $\ln \mathcal{L}_\mathrm{tot}$. Note also that the expressions for $\ln \mathcal{L}_\mathrm{HRS}$ and $\ln \mathcal{L}_\mathrm{LRS}$ are not the same, as the HRS likelihood is derived through a maximal likelihood estimator \citep[see][]{gibson2020} given that the derived uncertainties often underestimate the true uncertainty in the data because of uncorrected tellurics and/or systematics.

\section{Atmospheric Retrieval}\label{sec:retrieval}

We perform atmospheric retrievals of the datasets using the HyDRA-H retrieval framework \citep{gandhi2019_hydrah, gandhi2022}. These retrievals encompass the low-resolution (HST+Spitzer only) data, the high-resolution (CARMENES only) data and a combined retrieval across all datasets simultaneously. We determine the opacity arising from the spectrally active species in the atmosphere using molecular line lists from ExoMol for H$_2$O \citep{tennyson2016, polyansky2018} and HITEMP for CO and OH \citep{rothman2010, li2015}. We adopt H$_2$ and He broadened pressure broadening coefficients for H$_2$O and CO \citep[see][]{gandhi2020_cs}, but use air broadening for OH given that H$_2$/He broadening is not readily available. We also include the opacity from H- \citep{bell1987, john1988}, arising from the contribution due to bound-free and free-free transitions \citep[see][for further details on the cross section calculation of H-]{gandhi2020_h-}. We leave the chemical abundance of each species as a free parameter in our retrieval, resulting in four free parameters for the atmospheric chemistry. In addition to molecular line opacity we also include absorption from H$_2$-H$_2$ and H$_2$-He interactions as a source of continuum opacity \citep{richard2012}.

The solar abundance weighted cross section is given in Figure~\ref{fig:cs}. This shows the cross section of each species at 2900~K and 1~mbar, weighted by the volume mixing ratio of each species in chemical equilibrium. This shows that in the H-band range, the H$_2$O and OH have prominent opacity, with weaker contributions also arising from CO and H-. The CO is more prominent over the Spitzer channels, in particular the 4.5~$\mu$m range. We also show the photosphere (optical depth = 0.56 surface) for each of these species assuming thermal dissociation and an isothermal temperature profile of 2900~K at solar composition \citep{asplund2021}. The OH lines have the highest altitude photosphere, with some lines reaching pressures as low as 0.4~mbar, but an average at $\sim$1-2~mbar. On the other hand, the H$_2$O has generally weaker features but many more lines across the whole of the H band range, with typical photosphere pressures of $\sim$3-4~mbar. When multiple species are present in the atmosphere, the photosphere will generally trace the peak of the lines of each species, as the contribution to the opacity from each chemical species is summed in the calculation of the optical depth.

We determine the temperature profile using the \citet{madhu2009} parametrisation, with 6 free parameters which allow for a fully flexible thermal profile of the atmosphere capable of capturing non-inverted, isothermal and inverted temperature profiles. The prior range for each of these parameters is shown in Table~\ref{tab:priors}.
We also include the reference pressure for the measured planetary radius of 1.854~R$_\mathrm{J}$ as a retrieval parameter, in accordance with previous work \citep[e.g.,][]{welbanks2019_degen}, and a partial cloud/haze model with 4 free parameters \citep[see e.g.,][]{line2016, macdonald2017}. Hence we have 15 free parameters in each retrieval that we perform. For retrievals including the high-resolution CARMENES data we include 3 additional parameters, the deviation from the known systemic velocity ($\Delta V_\mathrm{sys}$), the deviation from the planet's known orbital velocity ($\Delta K_\mathrm{p}$), and the FWHM (full width half maximum) of the broadening kernel ($\delta V_\mathrm{wind}$) applied to the spectrum. This broadening is applied in addition to the planet's rotational broadening, and arises due to atmospheric dynamics \citep[see e.g.,][]{gandhi2022}. Our parameter estimation is performed using the Nested Sampling algorithm MultiNest \citep{feroz2008, feroz2009, buchner2014}.

\begin{table}
    \centering
    \begin{tabular}{c|c|c}
& \textbf{Parameter}              & \textbf{Prior Range}\\
\hline
Chemistry   & $\log(X_\mathrm{H_2O})$  & -12 $\rightarrow$ -1 \\
            & $\log(X_\mathrm{OH})$ & -12 $\rightarrow$ -1 \\
            & $\log(X_\mathrm{CO})$ & -12 $\rightarrow$ -1 \\
            & $\log(X_\mathrm{H-})$ & -12 $\rightarrow$ -1 \\
\hline
Temperature Profile & $T_\mathrm{0.1mb}$ / K & 1500 $\rightarrow$ 3800 \\
            & $\alpha_1\, /\, \mathrm{K}^{-\frac{1}{2}}$ & 0 $\rightarrow$ 1 \\
            & $\alpha_2\, /\, \mathrm{K}^{-\frac{1}{2}}$ & 0 $\rightarrow$ 1 \\
            & $\log(P_1 / \mathrm{bar})$ & -7 $\rightarrow$ 2 \\   
            & $\log(P_2 / \mathrm{bar})$ & -7 $\rightarrow$ 2 \\
            & $\log(P_3 / \mathrm{bar})$ & -2 $\rightarrow$ 2 \\
\hline
Ref. pressure & $\log(P_\mathrm{ref} / \mathrm{bar})$ & -4 $\rightarrow$ 2 \\
\hline
Clouds/hazes & $\log(\alpha_\mathrm{haze})$ & -4 $\rightarrow$ 6 \\
            & $\gamma_\mathrm{haze}$ & -20 $\rightarrow$ -1 \\
            & $\log(P_\mathrm{cl}/\mathrm{bar})$ & -7 $\rightarrow$ 2 \\
            & $\phi_\mathrm{cl}$ & 0 $\rightarrow$ 1 \\ 
\hline
HRS params  & $\Delta K_\mathrm{p}$ / kms$^{-1}$ &-80 $\rightarrow$ 80 \\
            & $\Delta V_\mathrm{sys}$ / kms$^{-1}$ & -30 $\rightarrow$ 30 \\
            & $\delta V_\mathrm{wind}$ / kms$^{-1}$  & 1 $\rightarrow$ 30\\
\hline
    \end{tabular}
    \caption{Parameters and uniform prior ranges for our retrieval. We retrieve the H$_2$O, OH, H- and CO abundances, temperature profile, and partial cloud/haze parameters. Our temperature profile includes 6 free parameters and uses the prescription of \citet{madhu2009}, and our partial cloud/haze parametrisation includes 4 free parameters (see section~\ref{sec:retrieval}).}
    \label{tab:priors}
\end{table}

\subsection{H$_2$O dissociation model}\label{sec:dissoc}

The atmospheres of UHJs reach temperatures $\gtrsim$2000~K, where H$_2$O is expected to dissociate into OH and H at pressures $\lesssim10^{-1}$¬bar \citep{parmentier2018, arcangeli2018}. Hence it is instructive to define a new model that allows for such altitude dependent chemistry. Such an approach is useful as vertically varying chemistry has been shown to significantly alter the inferred C/O ratios for UHJs \citep{ramkumar2023}. We use the prescription of \citet{parmentier2018} to determine the H$_2$O abundance as a function of pressure in a thermally dissociated atmosphere, with the undissociated/deep atmosphere abundance left as a free parameter in our retrieval. Such a model has been used to characterise the atmospheres of other UHJs \citep[e.g.][]{coulombe2023}. The volume mixing ratio $\mathrm{X_{H_2O}}(\mathrm{P},\mathrm{T})$ is given by
\begin{align}
    \frac{1}{\mathrm{X_\mathrm{H_2O}^{0.5}}} &= \frac{1}{\mathrm{X_{0,H_2O}^{0.5}}} + \frac{1}{\mathrm{X_{d,H_2O}^{0.5}}},\label{eqn:dissoc}
\end{align}
where $\mathrm{X_{0,H_2O}}$ refers to the deep atmosphere abundance, and 
\begin{align}
    \log(\mathrm{X_{d,H_2O}}) &= 2.0 \log(\mathrm{P}) + 4.83\times 10^4/\mathrm{T} - 15.9,
\end{align}
for pressure P(bar) and temperature T(K) \citep{parmentier2018}. This has a strong dependency on the pressure, with lower pressures resulting in substantial dissociation of H$_2$O. Between pressures of 5~mbar and 1~mbar, the H$_2$O abundance decreases by $\sim$1 dex at a temperature of 2900~K. As HRS observations are generally more sensitive to higher altitudes in the atmosphere than LRS \citep{gandhi2020_clouds}, we are sensitive to such vertical chemical gradients of the H$_2$O in the atmosphere when combining the observations. We leave the OH abundance as vertically-homogeneous for our retrievals, to ensure that our retrievals only explore the dissociation of the H$_2$O and the variation in its abundance. However, future work will extend this to OH, as OH is also expected to dissociate into O and H in the photosphere at high temperatures \citep[e.g.,][]{landman2021}. As before, there are two additional chemical parameters for the chemical abundances of CO and H-, which are also fixed with height in the atmosphere. Therefore, we perform 6 retrievals in total, using the LRS, HRS, combined (LRS+HRS) data, with both the H$_2$O-dissociation and vertically-homogeneous models.

\begin{figure*}
\centering
	\includegraphics[width=\textwidth,trim={0cm 0cm 0cm 0},clip]{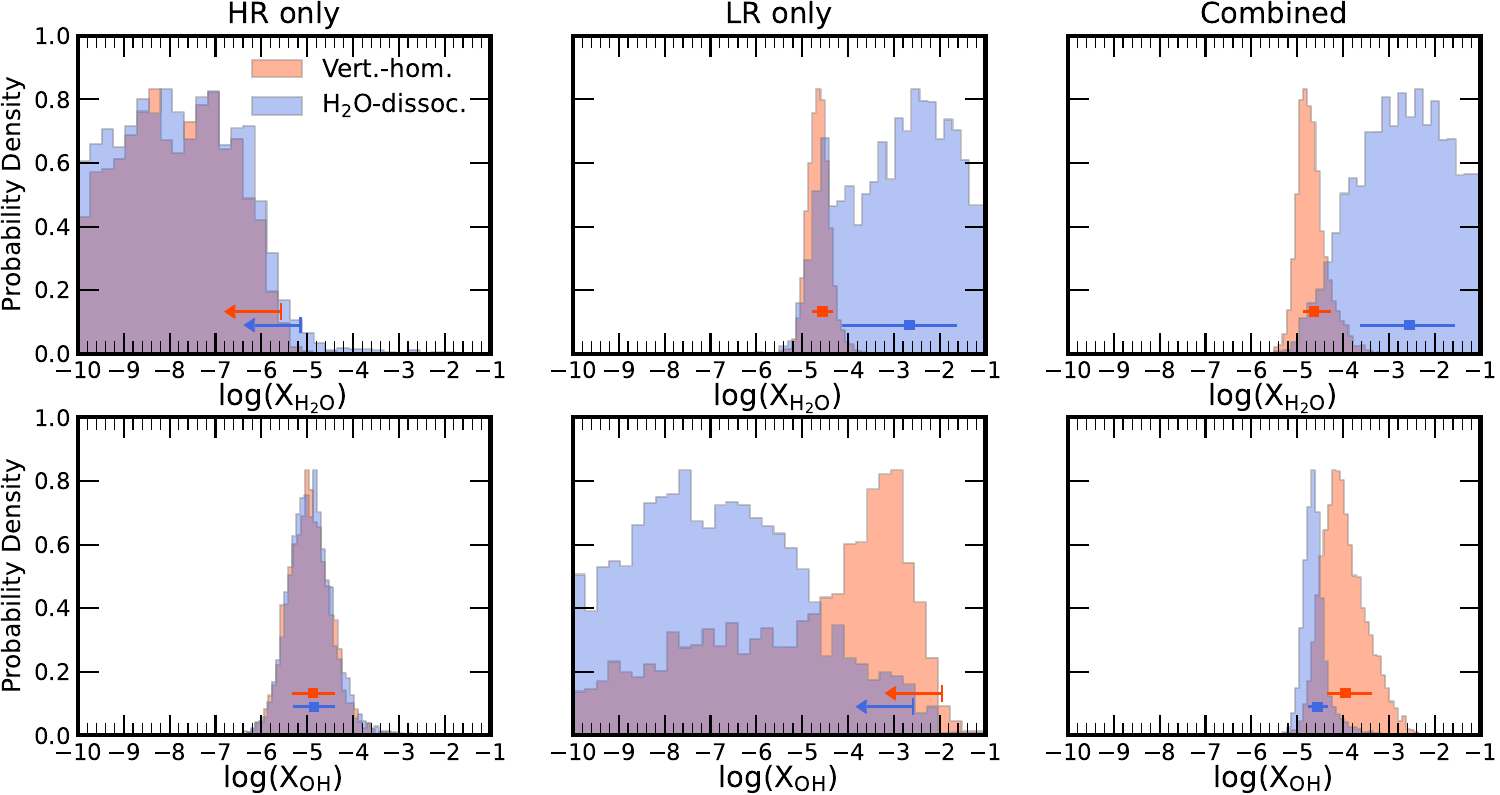}
    \caption{Retrieved volume mixing ratios for H$_2$O (top row) and OH (bottom row), performed with the high-resolution data only (left column), low-resolution data only (middle column), and combined datasets (right column). We show the constraints for the dissociation model in blue (for the deep atmosphere/undissociated abundance), and the H$_2$O-dissociation model in red. We also show the median value and $\pm1\sigma$ constraint for the cases where a species was detected, and 2$\sigma$ upper limits where no clear detection was obtained. The retrieved values and detection significances for each species are given in Table~\ref{tab:detection_significances}.}     
\label{fig:6panel}
\end{figure*}

\begin{table*}
    \centering
    \def\arraystretch{1.5}
    \begin{tabular}{cc|cc|cc}
        && \multicolumn{2}{c}{\textbf{H$_2$O}} & \multicolumn{2}{c}{\textbf{OH}}\\
        \textbf{Model} & \textbf{Retrieval} & log(X$_\mathrm{H_2O}$) & Det. Significance & log(X$_\mathrm{OH}$) & Det. Significance\\
        \hline
\textbf{Vertically-homogeneous H$_2$O}  & High-resolution & $<-5.6$ & - & $-4.9^{+0.5}_{-0.4}$ & 4.9$\sigma$\\
 & Low-resolution & $-4.6^{+0.2}_{-0.2}$ & 3.9$\sigma$ & $<-2.0$ & -\\
 & \textbf{Combined} & $-4.6^{+0.4}_{-0.2}$ & 4.0$\sigma$ & $-3.9^{+0.6}_{-0.4}$ & 4.5$\sigma$\\
\hline
\textbf{H$_2$O-dissociation}  & High-resolution & $<-5.1$ & - & $-4.9^{+0.5}_{-0.4}$ & 4.8$\sigma$\\
 & Low-resolution & $-2.7^{+1.0}_{-1.5}$ & 4.2$\sigma$ & $<-2.6$ & -\\
 & \textbf{Combined} & $-2.5^{+1.0}_{-1.1}$ & 4.4$\sigma$ & $-4.6^{+0.2}_{-0.2}$ & 4.3$\sigma$\\
    \end{tabular}
    \caption{Constrained volume mixing ratios and detection significances for H$_2$O and OH for each of the retrievals conducted.}
    \label{tab:detection_significances}
\end{table*}

\section{Results and discussion}\label{sec:results}

\begin{figure*}
\centering
	\includegraphics[width=\textwidth,trim={0cm 0cm 0cm 0},clip]{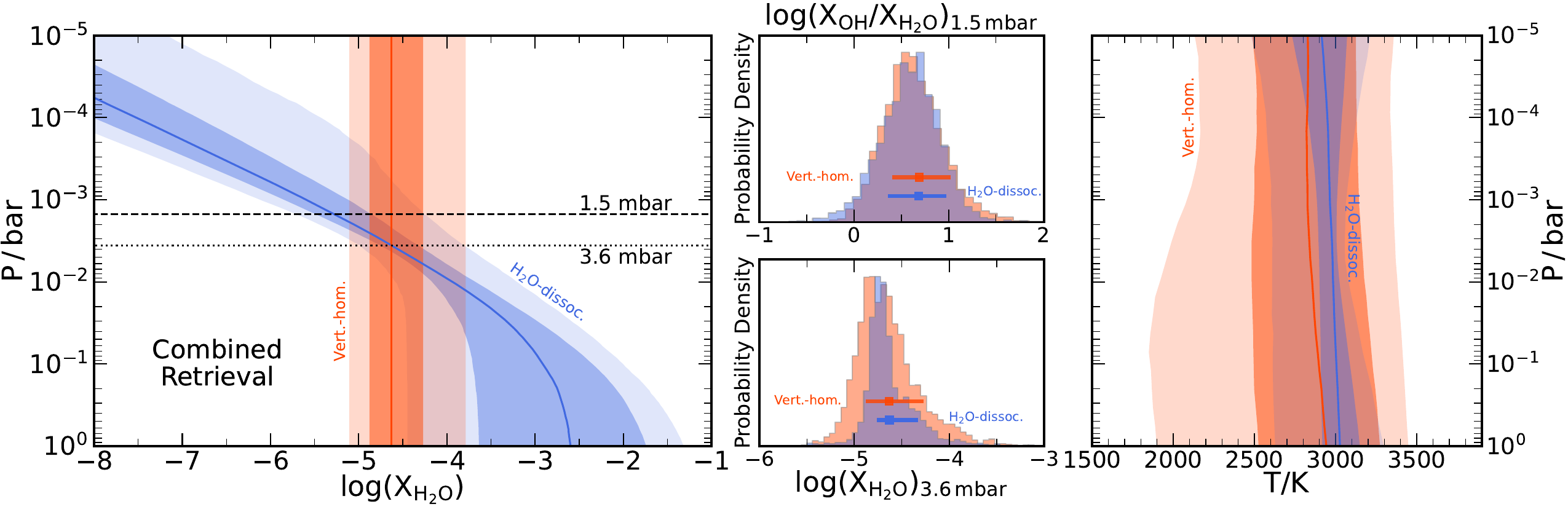}
    \caption{Left panel: Constraints on the chemical abundance profiles as a function of pressure for the H$_2$O-dissociation (blue) and vertically-homogeneous (orange) models for the combined retrievals. The top middle panel shows the constrained OH/H$_2$O ratio at 1.5~mbar, and the bottom middle panel shows the volume mixing ratio of H$_2$O for the WFC3 photosphere at 3.6~mbar. Right panel: Constraints on the temperature profile from both models.}
\label{fig:abundances}
\end{figure*}

In this section we discuss the results of our retrievals. We perform six retrievals in our study, using both H$_2$O-dissociation and vertically-homogeneous H$_2$O model across the HRS, LRS and combined datasets. Figure~\ref{fig:6panel} shows the abundance constraints for H$_2$O and OH for each of the retrievals. For the H$_2$O-dissociation model we show the deep atmosphere abundance of H$_2$O ($\mathrm{X_{0,H_2O}}$). We clearly constrain the OH using the high-resolution CARMENES observations, which show posterior peaks for both models. However, the H$_2$O abundance is not well constrained with the HRS data, with a slightly higher upper limit for the dissociation model. This is likely because the lower pressures probed by the HRS data are too depleted in H$_2$O to provide a clear constraint on its abundance. Furthermore, our high-resolution analysis is more optimised towards the OH, so the H$_2$O may not be clearly detected due to the data cleaning and model reprocessing removing much of the H$_2$O signal (see section~\ref{sec:hr_analysis}). The strong telluric absorption therefore makes constraining H$_2$O from the ground-based high-resolution spectroscopy more challenging \citep[e.g.,][]{brogi2012, brogi2019,pelletier2021}. In addition, the OH signal was detected with a strong velocity shift, as reported in \citet{landman2021} and \citet{cheverall2023}. If the H$_2$O spectral lines are shifted with respect to the OH lines, such as if there is any substantial chemical asymmetry or wind-shear in the atmosphere of the planet \citep[e.g.,][]{rauscher2014, wardenier2021}, the retrieval will converge to the position of the OH lines as that is the dominant signal in the data. \citet{wardenier2023} showed that for WASP-76~b, the OH should be more prominent on the evening/trailing terminator and would thus have a velocity shift relative to the H$_2$O which would be more prevalent on the morning/leading terminator. We tested this by performing retrievals with the HRS data that did not retrieve the OH and these did show a weak peak for H$_2$O, but at much lower $K_\mathrm{p}$ values than the OH signal or the expected velocity for the planet, and the overall evidence for H$_2$O was not significant to make robust assertions.

On the other hand, the LRS data clearly constrains the H$_2$O. The H$_2$O-dissociation model shows a higher constrained value with a much wider uncertainty as it is only able to capture a lower limit on the deep atmosphere H$_2$O given its detection in the photosphere. Our results are consistent with constraints from \citet{fu2021}, who retrieved H$_2$O at $\sim$1.1~dex sub-solar values. Given the weak broad band absorption of OH in the WFC3 range, the OH is unconstrained in both models in this case. Neither the HRS or LRS retrievals are able to provide any meaningful constraints on the H- given its lack of significant spectral features.

The combined retrievals, performed on the LRS and HRS data simultaneously, are able to constrain both H$_2$O and OH simultaneously, with detection significances $\gtrsim$4$\sigma$ for both species in the H$_2$O-dissociation and vertically-homogeneous H$_2$O models. The H$_2$O abundance remains identical to the LRS only case (see Table~\ref{tab:detection_significances}), indicating there is little information from the HRS data on the H$_2$O. The OH abundance does show a slight increase in abundance for the vertically-homogeneous model, but this change is still within 1$\sigma$ of the HRS only retrieval. Hence the combined retrievals allow for the most complete picture of the dissociating atmosphere of WASP-76~b, as either dataset alone is not able to pick up both species clearly.

The H$_2$O abundance in the WFC3 photosphere and the OH/H$_2$O ratio in the CARMENES photosphere are the same for both the H$_2$O-dissociation and vertically-homogeneous model (see Figure~\ref{fig:abundances}). The LRS data is most sensitive to the H$_2$O abundance at $\sim$3-4~mbar, as shown in Figure~\ref{fig:cs}, and thus it is unsurprising that the two models have an equal abundance at these pressures. On the other hand, abundance constraints with high-resolution spectroscopy are typically set by the spectral features over the continuum \citep[e.g.][]{gibson2022, maguire2023, pelletier2023}, which is primarily the H$_2$O opacity in this case. Therefore, the ratio of OH to H$_2$O is the quantity that remains unchanged across both models in the CARMENES photosphere at $\sim$1.5~mbar. At such pressures, the H$_2$O-dissociation model has a lower H$_2$O abundance than the vertically-homogeneous model given that $\mathrm{H_2O} \propto \mathrm{P}^2$ at pressures $\lesssim10^{-2}$~bar. Therefore, the absolute value of the constrained OH abundance lowers for the dissociation model to preserve this ratio between the two species. Given we are most sensitive to these quantities and they remain the same for both models, it is unsurprising that the two models have an almost equal Bayesian evidence in the combined retrievals, with only a slight preference for the H$_2$O-dissociation model at 1.9$\sigma$.

\begin{figure*}
\centering
	\includegraphics[width=\textwidth,trim={0cm 0cm 0cm 0},clip]{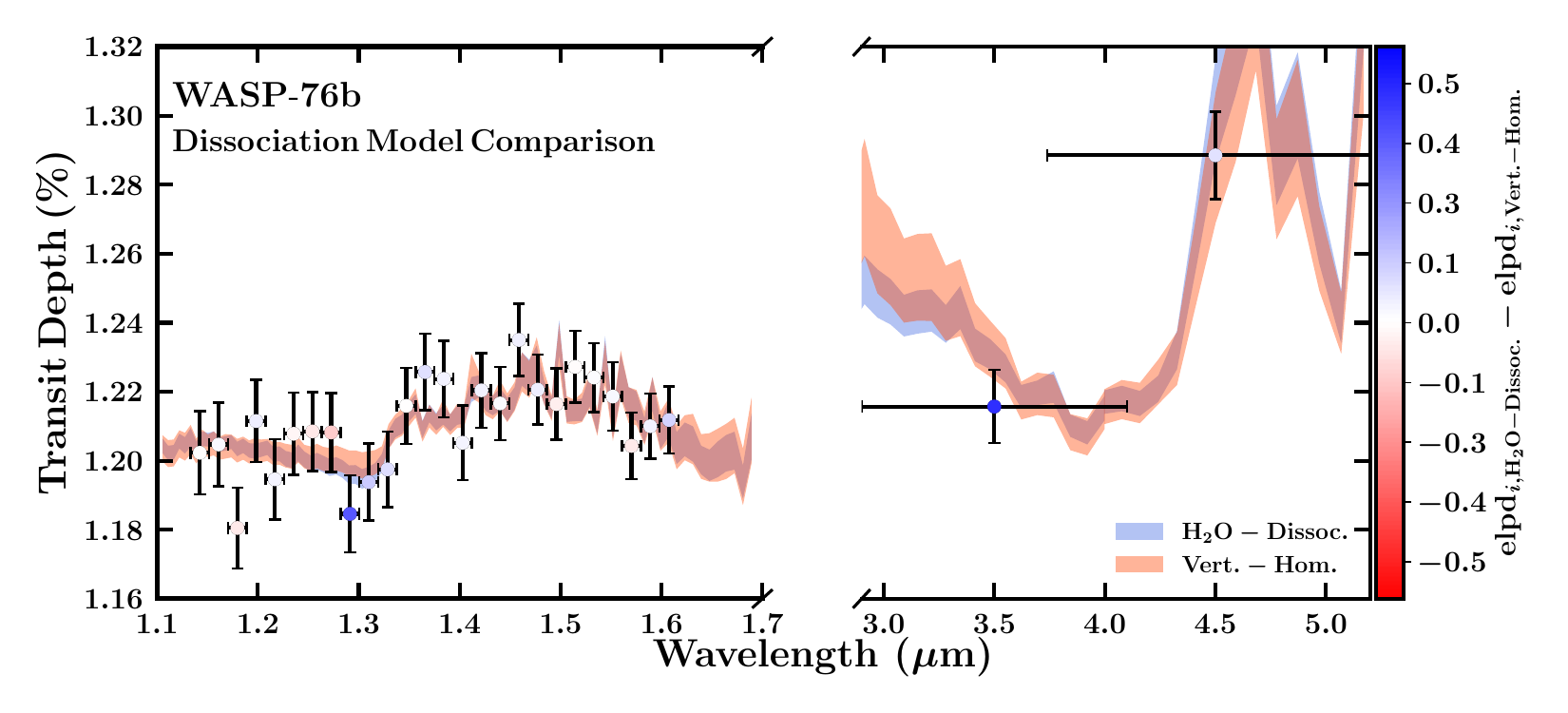}
    \caption{LOO-CV analysis of HST WFC3 and Spitzer spectro-photometric data from a joint retrieval with CARMENES observations. Data points are color-coded by their $\Delta$elpd$_{\text{LOO}}$ score between the model with H$_2$O-dissociation and the vertically-homogeneous H$_2$O model. Points with a larger $\Delta$elpd$_{\text{LOO}}$ score (bluer points) are best explained by the model including H$_2$O-dissociation, while points with a lower score (redder points) are best explained by the model without the dissociation. The shaded regions indicate the 1$\sigma$ retrieved models for the dissociation (blue) and vertically-homogeneous (orange) configurations.}
\label{fig:LOO}
\end{figure*}

For the HRS photosphere at 1.5~mbar, the ratio of abundances provide a direct constraint on the degree to which H$_2$O is dissociated. The OH/H$_2$O ratio is $\log(\mathrm{OH/H_2O}) = 0.7^{+0.3}_{-0.3}$ for both the H$_2$O-dissociation and vertically-homogeneous models. With the assumption that the OH is formed through the dissociation of H$_2$O, both models indicate that the majority of the H$_2$O is dissociated at 1.5~mbar given that the OH abundance exceeds the H$_2$O. The H$_2$O-dissociation model discussed in section~\ref{sec:dissoc} predicts that at a temperature of 2900~K and 1.5~mbar, the H$_2$O abundance is $\log(\mathrm{X_{H_2O}}) = -5.0$ for a solar composition atmosphere. This indicates that $\sim$98\% of the H$_2$O in the photosphere should be dissociated. If we assume that the only dissociation product of H$_2$O is OH, our retrievals indicate that 83$^{+8}_{-11}$\% of the H$_2$O is dissociated at 1.5~mbar, slightly lower than expected. However, we note that the value that our retrievals constrain is only a lower bound, as the OH will further dissociate to O and H \citep[e.g.,][]{landman2021, brogi2023}. Nevertheless, the constraints on H$_2$O and OH from the combined retrievals are able to provide direct evidence of significant thermal dissociation of H$_2$O in UHJs regardless of model choice. We performed retrievals assuming the OH abundance was set by subtracting the dissociated H$_2$O abundance from the undissociated/deep atmosphere H$_2$O, but this resulted in unphysical abundance constraints and was statistically disfavoured by $\gtrsim3\sigma$. This is because the OH is also substantially dissociated in the photosphere and hence the sum of the OH and H$_2$O is significantly less than the total oxygen content of the atmosphere. Hence the assumption that the dissociated H$_2$O equates to the OH abundance is not valid for UHJ atmospheres.

In addition, the temperature profile will strongly affect the degree of dissociation. Our P-T profiles are shown in the right hand panel of Figure~\ref{fig:abundances}, and indicate that the P-T profile is better constrained for the H$_2$O-dissociation model. This is because this model couples the atmospheric abundance of H$_2$O to the temperature, as the degree of dissociation of the H$_2$O is directly affected by the temperature profile. On the other hand, our vertically-homogeneous model allows for a much wider range of temperatures as these are largely just set by the scale height of the atmosphere. The median best-fit profile and 1$\sigma$ confidence intervals are however consistent across both models for the whole pressure range of the model and similar to the equilibrium temperature of WASP-76~b, which is encouraging.

We further explored the differences between the two models using Bayesian Leave-one-out Cross-validation (LOO-CV) for exoplanet atmospheres as introduced in \citet{Welbanks2023}. Unlike the Bayesian Evidence which provides a single value shedding little light on exactly where and how each model is performing well or poorly, LOO-CV enables us to calculate the expected log pointwise predictive density (elpd$_{\text{LOO}}$) and assess model performance at the per-data-point level. Here, we calculate the elpd$_{\text{LOO}}$ for each data-point, for each of the H$_2$O-dissociation and vertically-homogeneous H$_2$O models, from the posterior samples of each of the retrievals using the Pareto Smoothed Importance Sampling (PSIS) approximation from \citet[][]{Vehtari2015, Vehtari2017} as explained in \citet{Welbanks2023}, with the exception of the Spitzer 4.5$\mu$m channel for which a full Bayesian refit with the data point left out was necessary. 

\begin{figure*}
\centering
	\includegraphics[width=\textwidth,trim={0cm 0cm 0cm 0},clip]{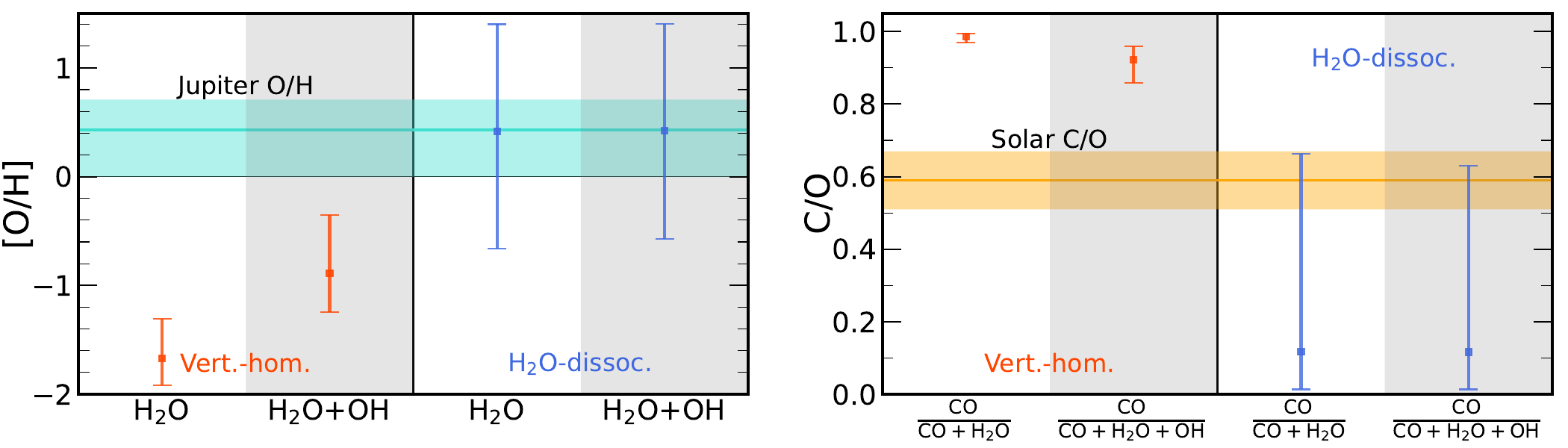}
    \caption{Left panel: Inferred bulk [O/H] values from the combined retrievals with the H$_2$O-dissociation and vertically-homogeneous models. We show the constraints from H$_2$O+OH together as well as H$_2$O separately. We also show the [O/H] for Jupiter \citep{li2020}. Right panel: Derived C/O ratios for each of the models, along with the solar C/O \citep{asplund2021}.}
\label{fig:ratio}
\end{figure*}

Figure \ref{fig:LOO} presents the comparison between the vertically-homogeneous and H$_2$O-dissociation models using the elpd$_{\text{LOO}}$ score for the HST and Spitzer observations. Our LOO-CV analysis finds that the two best-explained data points by the H$_2$O-dissociation model are the Spitzer 3.5 $\mu$m channel followed by the HST-WFC3 wavelength bin centered at $\sim1.29\mu$m. Indeed, the dissociation model offers the flexibility to have lower transit depths near those two data, compared to the no dissociation model, as shown by the 1$\sigma$ model contours shown in Figure \ref{fig:LOO}. The predictive model performance as quantified by the sum of the $\Delta$elpd$_{\text{LOO}}$ scores\citep[equation 6 in][]{Welbanks2023} finds that the inclusion of H$_2$O dissociation increases the predictive performance of the models by 1.8 standard errors (e.g., $\Delta$elpd$_{\text{LOO}}$/SE, $\Delta$elpd$_{\text{LOO}}$ = 2.5, SE = 1.8).

We confirm the intuition provided by the LOO-CV analysis above by performing retrievals using both H$_2$O-dissociation and vertically-homogeneous models but not considering the 3.5$\mu$m and $1.29\mu$m observations. For this scenario, the marginal preference for the H$_2$O-dissociation model as suggested by the Bayesian evidence comparison is removed. This insight suggests that additional observations in the infrared can provide the required information to assess the current marginal preference for the H$_2$O-dissociation model. Since removing two low-resolution data removes the marginal preference for the H$_2$O-dissociation model in the combined LR+HR retrievals, we confirm that the CARMENES observations do not have a strong preference for either model. This is because the H$_2$O is not clearly detected from the high-resolution data, and the two models' Bayesian evidences are thus very similar. Drawing a preference for the H$_2$O-dissociation model based alone on the interpretation of model comparison metrics such as the Bayesian Evidence or elpd$_{\text{LOO}}$ will therefore require joint analysis with upcoming observations with JWST and ground-based facilities. 

The H$_2$O-dissociation and vertically-homogeneous models do significantly differ in their inferred bulk oxygen content of WASP-76~b, as shown in Figure~\ref{fig:ratio}. At high pressure, the majority of the H$_2$O is undissociated, and thus the two models diverge in their deep atmosphere constraints (see Figure~\ref{fig:abundances}). The dissociation model offers little more than a lower bound on the H$_2$O and thus the O/H ratio, with the lower limit largely set by the value constrained at $\sim$mbar level. This O/H value is consistent with the O/H for Jupiter, but has significantly larger error bars which allow for a wide range of possible solutions from sub-Jovian to super-Jovian values. The vertically-homogeneous model on the other hand gives a much lower O/H abundance, set through the photosphere constraints. For this model, including the OH abundance to determine the O/H is key as the OH abundance is greater than the H$_2$O abundance and hence significantly affects the bulk O/H value. However, the 1$\sigma$ range for both the H$_2$O only and combined H$_2$O+OH constraint is lower than that for Jupiter given the low abundances constrained for the photosphere.

The degree of vertical mixing can also influence the photospheric H$_2$O abundance. Weak vertical mixing would result in the equilibrium processes and hence thermal dissociation dominating, and would result in low pressures having a substantial depletion of H$_2$O. On the other hand, if the vertical mixing dominates the chemical profile of the atmosphere would not have significant chemical gradients, more akin to the vertically-homogeneous model. Furthermore, given that we are probing the limb of the planet, some regions may probe the cooler non-irradiated regions of the atmosphere and some more strongly irradiated, potentially resulting in patchy regions with differing abundances depending on the reaction timescales \citep{parmentier2018}. Therefore, understanding non-equilibrium processes is critical in obtaining accurate constraints on the bulk composition. 

The choice of model also has a knock on effect on the C/O ratio, with the inferred C/O ratio significantly lower for the H$_2$O-dissociation model given the higher bulk oxygen abundance (see Figure~\ref{fig:ratio}). Retrievals of other UHJs has showed that self-consistent models that incorporate dissociative processes lower the C/O ratio \citep{brogi2023, ramkumar2023}, as the constraints on the deep atmosphere abundances and hence C/O ratio vary significantly depending on the modelling assumptions. This was also seen in 3-dimensional general circulation models by \citet{pluriel2020}, who showed that a dissociating atmosphere influences the inferred CO/H$_2$O abundance. The C/O ratio and metallicity (or bulk O/H) are key parameters in formation \citep[e.g.][]{oberg2011, madhu2014, mordasini2016}, and understanding the influence of thermal dissociation is vital to constrain the formation history of exoplanets. We infer the C/O ratio by retrieving the CO abundance as a proxy for the carbon in the atmosphere, which is constrained through the Spitzer 3.6~$\mu$m and 4.5~$\mu$m channels. The CO is a relatively stable molecule that is not expected to dissociate in the atmospheres of UHJs \citep[e.g.][]{parmentier2018}. The atmospheric C/O ratio can strongly impact the equilibrium H$_2$O abundance, with higher values of C/O depleting the H$_2$O in the atmosphere as the oxygen preferentially binds to carbon to form CO at high temperature \citep[e.g.][]{madhu2012}. Such a solution is therefore degenerate with thermal dissociation, as both deplete the photospheric H$_2$O. In addition, oxygen rich species such as CaTiO$_3$ may rain-out on the cooler night sides of UHJs \citep{hoeijmakers2023}, resulting in a loss of some of the oxygen which would subsequently drive the atmospheric C/O ratio to higher values.

Other carbon-bearing species may also be present and affect the C/O ratio of the atmosphere. We performed retrievals with HCN and CO$_2$, using molecular cross sections derived from the ExoMol \citep{harris2006, barber2014} and Ames \citep{huang2013, huang2017} line lists respectively, but found no evidence for either species. However, CO is expected to remain the dominant carbon source of the atmosphere and hence other species are unlikely to significantly alter the C/O ratio. Furthermore, \citet{sanchezlopez2022} indicated that chemical gradients may be present between the terminator regions from the CARMENES observations of WASP-76~b, and 3D simulations with JWST have shown that strong day-night temperature gradients can also impact the abundance constraints for UHJs \citep{pluriel2020}. At high resolution, chemical gradients/asymmetries may manifest as velocity shifts between species given the rotation and atmospheric dynamics, and WASP-76~b shows such asymmetries from observations \citep{demangeon2024}. \citet{sanchezlopez2022} indicated that there were signs of HCN in WASP-76~b, but at a significant velocity shift to the OH signal, and hence our retrieval may not be able to constrain it given the model setup does not allow velocity offsets between species. However, \citet{baeyens2023} showed that photochemistry can generate significant HCN on the morning terminator on UHJs. Therefore, further works with multi-dimensional retrievals \citep[e.g.,][]{gandhi2022,maguire2024} may be able to discern these if the signal-to-noise of the observations is sufficient.

\section{Conclusions}

In this work we performed retrievals combining ground-based high-resolution CARMENES observations with space-based low-resolution HST WFC3 and Spitzer observations of the primary eclipse of the ultra-hot Jupiter WASP-76~b. The HRS and LRS observations provide an excellent opportunity to test the capabilities of both methodologies, particularly given that the WFC3 and CARMENES wavelength ranges overlap in the $\sim$1.1-1.7~$\mu$m region. With spectral absorption for both H$_2$O and OH in the wavelength ranges probed by the observations, such combined retrievals present an ideal opportunity to explore these processes in giant exoplanets. We constrain OH from the high-resolution observations given that it possesses strong spectral lines in the H-band. On the other hand, we constrain H$_2$O primarily from the HST observations given the difficulty in constraining H$_2$O over strong telluric absorption with HRS and that H$_2$O may have velocity shifts relative to OH. Hence it is through combined retrievals that we are able to obtain a full picture of the dissociation of H$_2$O in WASP-76~b, highlighting the need for observations with both approaches in order to accurately obtain the total oxygen content of the atmosphere.

Given the high equilibrium temperature of UHJs such as WASP-76~b, H$_2$O is expected to be thermally dissociated into OH and H in the photosphere. To explore the dissociation of H$_2$O we adopted a dissociation model into our retrievals and compare our retrieved abundance constraints and overall Bayesian evidences with a model with a vertically-homogeneous chemical profile. Model comparison using the Bayesian evidence finds marginal preference for the H$_2$O-dissociation model. Similarly, a LOO-CV analysis finds that the inclusion H$_2$O-dissociation results in only a slight increase to the predictive performance of the model. Furthermore, our LOO-CV analysis highlights that the photometric 3.5$\mu$m channel is key in driving this marginal model preference. This is because the two quantities most sensitive in our observations and which remain the same across both models are the H$_2$O abundance in the HST WFC3 photosphere ($\sim$3-4~mbar), and the OH/H$_2$O ratio in the CARMENES photosphere ($\sim$1-2~mbar). The low-resolution observations generally probe the continuum and hence are more sensitive to the absolute value of the H$_2$O abundance at the slightly deeper $\sim$3-4~mbar level. However, HRS is more sensitive to the relative line positions of the OH lines over the continuum and generally probes lower pressures, and hence it is the ratio of the two species is consistent at 1.5~mbar. Hence, photospheric abundance ratios constrained from retrievals of high resolution spectra are likely to be significantly more robust than absolute values, as found in previous studies \citep[e.g.][]{gibson2022, maguire2023, pelletier2023}. We find that for both models the OH abundance at 1.5~mbar is $\sim$0.7~dex greater than the H$_2$O, indicating that the majority of the H$_2$O in the photosphere has dissociated. Note that the OH further dissociates into its constituent atoms at higher temperatures/lower pressures so this is only a lower limit on the degree of H$_2$O dissociation.

Despite the similarity of the photospheric constraints, the bulk oxygen composition inferred between the H$_2$O-dissociation and vertically-homogeneous models differs significantly. The vertically varying abundance of the H$_2$O-dissociation model means the deep atmosphere H$_2$O is significantly greater and sets little more than a lower limit on the bulk O/H ratio. This has a knock on effect on the atmospheric C/O ratio, with the dissociation model indicating a lower ratio, similar to previous work using chemical equilibrium models \citep{brogi2023, ramkumar2023}. Therefore, we highlight the importance of the model choice in determining the bulk composition of exoplanet atmospheres in the presence of strong chemical gradients.

Combined spectroscopy offers a powerful tool in the characterisation of exoplanet atmospheres. Analysis with HRS and LRS data simultaneously allows us to combine the strengths of each method and leads to robust detections of chemical species. In particular, numerous optical spectrographs have constrained refractory species with strong opacity in the visible, and JWST has opened new avenues in atmospheric characterisation of volatile species in the infrared \citep[e.g.][]{JTEC2023, fu2022, constantinou2023}. In future, combined observations across optical and infrared wavelengths will be key for refractory-to-volatile abundance constraints \citep{lothringer2021, kasper2022}. In addition, JWST's MIRI has the capability to explore beyond 10$\mu$m, thereby opening up longer wavelengths where ground based observations struggle due to thermal background, and which are currently limited to observations no redder than the L-band \citep{hawker2018, webb2020}. Additionally, JWST NIRSPEC and JWST NIRCAM observations covering the $\sim 2\mu$m to $\sim 5\mu$m range \citep[e.g.,][]{Ahrer2023, Alderson2023} will be key to distinguish and constrain the effects of H$_2$O dissociation as highlighted by our LOO-CV analysis. Finally, with its ability to observe multiple water features, JWST NIRISS \citep[e.g.,][]{Feinstein2023, coulombe2023} may enable more robust constraints on the abundance of this species. Overall, future analysis combining JWST observations with high-resolution ground based observations \citep[e.g.,][]{august2023,smith2024} can help derive a better understanding of the chemical processes in WASP-76b. A possible extension to this work would be to introduce a two-layer parametrization to infer differences in chemical constraints between different pressure regions of the atmosphere \citep[e.g.,][]{changeat2019}, particularly given that the two sets of observations are sensitive to different altitudes. In the near future, combined spectroscopy with ELT and JWST will maximise the potential of these flagship missions, particularly for cooler, fainter and potentially more Earth-like exoplanets.

\section*{Acknowledgements}

SG is grateful to Leiden Observatory at Leiden University for the award of the Oort Fellowship. This work was performed using the compute resources from the Academic Leiden Interdisciplinary Cluster Environment (ALICE) provided by Leiden University. We also utilise the Avon HPC cluster managed by the Scientific Computing Research Technology Platform (SCRTP) at the University of Warwick. We thank the anonymous referee for a careful review of our manuscript.

\section*{Data Availability}

The models underlying this article will be shared on reasonable request to the corresponding author.



\bibliographystyle{mnras}
\bibliography{refs} 




\bsp	
\label{lastpage}
\end{document}